\documentstyle[preprint,eqsecnum,aps]{revtex}

\preprint{RCNP-Th00018}

\newcommand{\be}{\begin{equation}}
\newcommand{\ee}{\end{equation}}
\newcommand{\bea}{\begin{eqnarray}}
\newcommand{\eea}{\end{eqnarray}}
\newcommand{\bvec}[1]{\mbox{\boldmath $#1$}}

\begin{document}
\draft

\title{Weyl Invariant Formulation of Flux-Tube Solution\\ 
in the Dual Ginzburg-Landau Theory}

\author{Yoshiaki Koma\footnote{Email address: koma@rcnp.osaka-u.ac.jp} 
and Hiroshi Toki}
\address{Research Center for Nuclear Physics (RCNP), Osaka University\\
Mihogaoka 10-1, Ibaraki, Osaka 567-0047, Japan}

\date{\today}
\maketitle

\begin{abstract}\baselineskip = 0.65cm
The flux-tube solution in the dual Ginzburg-Landau (DGL) theory 
in the Bogomol'nyi limit is studied by using the manifestly 
Weyl invariant form of the DGL Lagrangian.
The dual gauge symmetry is extended to $[U(1)]_m^3$, and accordingly,
there appear three different types of the flux-tube.
The string tension for each flux-tube is calculated analytically 
and is found to be the same owing to the Weyl symmetry.
It is suggested that the flux-tube can be treated in
quite a similar way with the Abrikosov-Nielsen-Olesen vortex 
in the $U(1)$ Abelian Higgs theory except for 
various types of flux-tube.\\
\end{abstract}


\baselineskip = 0.7cm


\section{Introduction}\label{sec:intro}
Recent studies in lattice QCD in the maximally
Abelian gauge suggest remarkable properties of the QCD vacuum,
such as Abelian dominance\cite{Brandstater:1991sn}
and monopole condensation\cite{Kronfeld:1987vd},
which provide the dual superconductor picture of the QCD vacuum
as is described by the dual Ginzburg-Landau (DGL) theory
\cite{Suzuki:1988yq,Suganuma:1995ps}.
The  DGL theory is obtained by using the 
Abelian projection\cite{'tHooft:1981ht}.
In this scheme, QCD is reduced into the $[U(1)]^2$ gauge theory 
including color-magnetic monopoles.
Based on the dual superconductor picture of the QCD vacuum, 
we get an intuitive picture of hadrons as 
the vortex excitation of the color-electric 
flux\cite{Nambu:1974zg,Mandelstam:1974pi}, 
which we call the color-electric flux-tube, or simply the flux-tube.
In this vacuum, the color-electric flux is squeezed into an almost 
one dimensional object like a string due to the dual Meissner effect caused 
by monopole condensation.
This situation seems to be the same with the appearance of the 
Abrikosov vortex in the ordinary superconductor system, which is 
caused by the Cooper pair condensation.

\par
We know that Abrikosov-Nielsen-Olesen (ANO) vortex 
in the ordinary superconductor can be described 
by using the Abelian Higgs theory\cite{Nielsen:1973ve}, 
where the keyword is the breaking of $U(1)_e$ gauge symmetry 
through the Higgs mechanism. 
Moreover, there exists an analytic solution of the 
ANO vortex in the border of the type-I and the type-II  vacuum, 
so called the Bogomol'nyi limit\cite{Bogomolny:1976de,deVega:1976mi}.
The analytical solution exhibits interesting features of the 
superconductivity and is useful to understand the properties 
of the vortex dynamics.
Hence, it is considered quite interesting to investigate the 
flux-tube solution in the dual superconductor QCD vacuum 
corresponding to the ANO vortex in the Bogomol'nyi limit.

\par
However, the symmetry in the QCD vacuum is not so simple compared with the 
ordinary superconductor system, since now we have to take into account the 
$[U(1)]_m^2$ dual gauge symmetry corresponding 
to the $U(1)_e$ gauge symmetry in the ordinary superconductor.
Note that the symmetry $[U(1)]^{N-1}$ is originated from 
the maximal torus subgroup of $SU(N)$.
Furthermore, we also have the {\em Weyl} symmetry, which is the 
permutation invariance of the labels among the Abelian color charges.
Therefore, the flux-tube in the QCD vacuum is expected to have 
some characteristic aspects beyond the analogue of the 
ANO vortex in the ordinary superconductor system.

\par
In this paper, we investigate the flux-tube solution in the DGL theory
in the Bogomol'nyi limit.
This study seems to be similar as is given in Ref. \cite{Chernodub:1999xi}. 
In fact, our result will be shown identical.
However, we would like to present an usuful method to find the
Bogomol'nyi limit, and this can be achieved by taking into 
account the Weyl symmetry in the DGL theory.
This idea can be extended straightforwardly 
to the $[U(1)]^{N-1}$ dual Abelian Higgs 
theory which would be reduced from the $SU(N)$ 
gluodynamics\cite{Ezawa:1981ub}.
We first write the DGL Lagrangian in a manifestly Weyl 
invariant form.
At the same time, we pay attention to the singular structure 
in the DGL theory, since it plays a significant role to obtain the 
string-like flux-tube solution.
Note that the boundary condition of the dual gauge 
field depends crucially on this singular structure.
Second, we consider the Bogomol'nyi limit, the border 
between the type-I and the type-II vacuum.
The string tension in this limit is computed analytically.
Finally, we discuss the properties of the flux-tube solution 
in the DGL theory.

\section{Manifestly Weyl invariant form of the DGL Lagrangian}\label{sec:Weyl}
The DGL Lagrangian\cite{Suzuki:1988yq,Suganuma:1995ps} 
is given by
\footnote{
\baselineskip = 0.6cm
Throughout this paper, we use the following notations:
Latin indices i,j express the labels 1,2,3, which is not
to be summed over unless explicitly stated.
Boldface letter denotes three-vector.}
\bea
{\cal L}_{\rm DGL} &=&
-\frac{1}{4} 
\left (
 \partial_{\mu}\vec{B}_{\nu}-\partial_{\nu}\vec{B}_{\mu} 
-\frac{1}{n \cdot \partial}\varepsilon_{\mu \nu \alpha \beta}
n^{\alpha} \vec{j}^{\beta} 
\right )^2 \nonumber\\
&&
  +\sum_{i=1}^3 
  \left [ \left | \left (\partial_{\mu}+ig\vec{\epsilon}
_{i}{\cdot}\vec{B}_{\mu} \right )\chi_{i} \right |^2
-\lambda \left ( \left |\chi_{i} \right |^2-v^2 \right )^2 \right ],
\label{eqn:DGL}
\eea
where $\vec{B}_{\mu}$ and $\chi_{i}$ denote the dual gauge field with 
two components $(B_{\mu}^3, B_{\mu}^8)$ and the complex scalar monopole 
field, respectively. 
The quark field is included in the current
$\vec{j}_{\mu} = e \bar{q}\gamma_{\mu} \vec{H} q$, 
$\vec{H}=(T_{3},T_{8})$.
Here, $\vec{\epsilon}_i$ is the root 
vector of $SU(3)$ algebra,
$\vec{\epsilon}_1=\left (-1/2,\sqrt{3}/2 \right ), 
\vec{\epsilon}_2=\left(-1/2,-\sqrt{3}/2 \right )
, \vec{\epsilon}_3=\left (1,0 \right )$, and $n^{\mu}$ denotes 
an arbitrary constant 
4-vector\footnote{\baselineskip = 0.6cm
If the dual gauge symmetry
is broken through monopole condensation, $n^{\mu}$ can not be an
arbitrary vector any more. Instead, this vector describes the 
dynamics of the string and gives the contribution
to the energy of the system.},
which corresponds to the direction of the Dirac string.
The gauge coupling $e$ and the dual gauge coupling $g$ 
hold the relation $eg=4\pi$.
This relation guarantees the unobservability of the Dirac string
when the dual gauge symmetry is not broken.
Note that the DGL Lagrangian (\ref{eqn:DGL}) is invariant under the 
$[U(1)]^2_{m}$ dual gauge transformation,
\bea
&&\chi_{i} \to  \chi_{i} e^{if_{i}}, \quad 
\chi^*_{i} \to \chi^*_{i} e^{-if_{i}},\nonumber\\
&&\vec{B}_{\mu} =(B_{\mu}^3,\; B_{\mu}^8)
\to
\left ( B_{\mu}^3 -\frac{1}{g}\partial_{\mu}f_3,\;
B_{\mu}^8 - \frac{1}{\sqrt{3}g} (\partial_{\mu}f_1 -\partial_{\mu}f_2)
\right ),\quad\quad (i=1,2,3),
\label{eqn:gauge-sym}
\eea
where the phase $f_i$ has the constraint
$\sum_{i=1}^3 f_{i}= 0$\cite{Suzuki:1988yq,Suganuma:1995ps}.

\par
The non-local term, in the kinetic term of the dual gauge field,
is concretely written as
\be
\frac{1}{n \cdot \partial} \varepsilon_{\mu \nu \alpha \beta}
n^{\alpha} \vec{j}^{\beta} 
=  
\int d^4 x' \langle x| \frac{1}{n \cdot \partial} |x' \rangle
\varepsilon_{\mu \nu \alpha \beta} n^{\alpha} \vec{j}^{\beta}(x'),
\ee
where
\bea
\langle x| \frac{1}{n \cdot \partial} |x' \rangle
&=&
\left [ p \theta ( (x- x')\cdot n )- (1-p) \theta ((x' - x)\cdot n )
\right ] \delta^{(3)} (\vec{x}_\perp - \vec{x}'_\perp ).
\eea
Here $p$ is an arbitrary real number and 
$\delta^{(3)}(x)$ is the $\delta$-function defined
on a three dimensional hyper-surface which has the normal vector $n_{\mu}$,
so that $\vec{x}_\perp$ and $\vec{x}'_\perp$ are 3-vectors 
(not necessarily spatial) which are perpendicular to the $n_{\mu}$.
It is noted that in order to define the color-electric charge of the quark 
in terms of the {\em dual} gauge field, we need such a non-local term, 
which is a result of the choice of 
{\em one potential approach}\cite{Blagojevic:1979bm}.

\par
Now, we define an extended dual gauge field to take into account the
Weyl invariance in the DGL theory as
\bea
B_{i\; \mu} \equiv 
\sqrt{\frac{2}{3}} \vec{\epsilon}_{i} \cdot \vec{B}_{\mu},
\quad\quad (i=1,2,3)
\label{eqn:b-weyl}
\eea
Here, the constraint $\sum_{i=1}^3 B_{i\; \mu} =0$ appears, which
has the same structure with the constraint $\sum_{i=1}^3 f_{i} =0$.
Furthermore, we divide the dual gauge field into two parts, the 
regular part and the singular part\cite{Ball:1988cf,Baker:1991bc},
\bea
\vec{B}_{\mu} \equiv \vec{B}_{\mu}^{\rm reg} +\vec{B}_{\mu}^{\rm sing}.
\eea
The factor $\sqrt{2/3}$ in (\ref{eqn:b-weyl}) is a simple normalization 
to get the factor $1/4$ in front of
the kinetic term of the dual gauge field (See (\ref{eqn:dgl-weyl})).
The singular dual gauge field $\vec{B}_{\mu}^{\rm sing}$ is determined
so as to cancel the Dirac string in the non-local term as
\bea
\partial_{\mu} \vec{B}_{\nu}^{\rm sing}-
\partial_{\nu} \vec{B}_{\mu}^{\rm sing} - 
\frac{1}{n \cdot \partial} \varepsilon_{\mu \nu \alpha \beta}
n^{\alpha} \vec{j}^{\beta} \equiv \vec{C}_{\mu \nu}.
\label{eqn:subtraction}
\eea
In the static $q$-$\bar{q}$ system, $\vec{C}_{\mu \nu}$ 
is nothing but the color-electric field originated from the 
color-electric charge like the electric field induced 
by an electric charge, where
an explicit form of $\vec{B}_{\mu}^{\rm sing}$ is given 
in Sec. \ref{sec:static}.
It is noted that the cross term of the regular dual field tensor
${}^* \! \vec{F}_{\mu \nu}^{\rm reg} \equiv
 \partial_{\mu} \vec{B}_{\nu}^{\rm reg}
-\partial_{\nu} \vec{B}_{\mu}^{\rm reg}$
and $\vec{C}_{\mu \nu}$ can be integrated out, and  
the square of $\vec{C}_{\mu \nu}$ and its integration 
gives the Coulomb energy including the self-energy of the color-electric 
charge. However, we drop it hearafter in order to concentrate
on the flux-tube itself. Correspondingly, we pay attention to the string 
tension for an ideal flux-tube system which has terminals at 
infinity\footnote{\baselineskip = 0.6cm
In order to classify the types of the flux-tube, 
we use the word such as the $q$-$\bar{q}$ system.}.

\par
Then, we obtain
\bea
{\cal L}_{\rm DGL} 
=
\sum_{i=1}^3 
\left [
- \frac{1}{4} {{}^* \! F_{i\;\mu \nu}^{\rm reg}}^2
+
\left | \left (\partial_{\mu} 
+ i g' 
\left ( 
B_{i\;\mu}^{\rm reg} + B_{i\;\mu}^{\rm sing}
\right )
\right )\chi_{i} \right |^2
- \lambda \left ( \left |\chi_{i} \right |^2-v^2 \right )^2 
\right ],
\label{eqn:dgl-weyl}
\eea
\bea
{}^* \! F_{i\;\mu \nu}^{\rm reg} \equiv
 \partial_{\mu} B_{i\;\nu}^{\rm reg}
-\partial_{\nu} B_{i\;\mu}^{\rm reg},
\eea
where the dual gauge coupling $g$ is scaled as
\be
g' \equiv \sqrt{\frac{3}{2}} g.
\ee
One finds that the dual gauge symmetry becomes very easy to observe, since
the dual gauge transformation is defined by
\bea
&&\chi_{i} \to  \chi_{i} e^{if_{i}},\quad
\chi^*_{i} \to \chi^*_{i} e^{-if_{i}},\nonumber\\
&&
B_{i\;\mu}^{\rm reg} \to
B_{i\;\mu}^{\rm reg} -\frac{1}{g'}\partial_{\mu}f_i, \quad\quad ( i=1,2,3)
\label{eqn:gauge-sym-weyl}
\eea
and accordingly the Lagrangian (\ref{eqn:dgl-weyl}) has the 
extended local symmetry $[U(1)]_m^3$. 
However, it does not mean an increase of the gauge degrees of freedom
because we have the constraint $\sum_{i=1}^3 B_{i\;\mu} = 0$.

\par
The field equations are given by
\bea
\left ( \partial_{\mu} +ig' 
\left ( B_{i\;\mu}^{\rm reg} + B_{i\;\mu}^{\rm sing} \right )
\right )^2 \chi_i
=
- 2 \lambda \chi_{i} ( \chi_{i}^{*} \chi_{i} - v^2),
\label{eqn:feq-1}
\eea
\bea
\partial^{\nu} {}^{* \!} F_{i\; \mu \nu}^{\rm reg} \equiv 
k_{i\;\mu} =
-ig' 
\left ( 
\chi_{i}^{*}\partial_{\mu} \chi_{i} - \chi_{i}\partial_{\mu} \chi_{i}^{*}
\right )
+
2 g'^2
\left ( 
B_{i\;\mu}^{\rm reg} + B_{i\;\mu}^{\rm sing}
\right ) \chi_{i}^{*}\chi_{i} ,
\label{eqn:feq-2}
\eea
These field equations are to be solved with 
the proper boundary conditions that quantize the color-electric 
flux\cite{Nielsen:1973ve}.
The flux is given by
\be
\Phi_i \equiv \int {}^{*} F_{i\;\mu \nu}^{\rm reg} d \sigma^{\mu \nu} =
\oint B_{i\;\mu}^{\rm reg} dx^{\mu},
\label{eqn:flux}
\ee
where $\sigma^{\mu \nu}$ is a two-dimensional surface element in the
Minkowski space.
By using the polar decomposition of the monopole field
as $\chi_i = \phi_i e^{i\eta_i}$ $(\phi_i, \eta_i \in \Re)$,
we get, from the field equation (\ref{eqn:feq-2}),
\bea
B_{i\;\mu}^{\rm reg} = \frac{k_{i\;\mu}}{2 {g'}^2 \phi_i^2}
-B_{i\;\mu}^{\rm sing}-\frac{1}{g'}\partial_{\mu}\eta_i .
\eea
We substitute this expression into (\ref{eqn:flux}) and
integrate out over a large closed loop where the monopole current 
$k_{i\;\mu}$ is vanished. Thus we get
\be
\Phi_i = - \oint \left ( B_{i\;\mu}^{\rm sing} 
+ \frac{1}{g'} \partial_{\mu}\eta_i 
\right ) dx^{\mu} .
\label{eqn:flux2}
\ee

\par
It is suggested from this expression that there are two 
possibilities to obtain the flux-tube configuration.
One is originated from the singularity in $B_{i\;\mu}^{\rm sing}$ and 
the other is from the singularity in $\partial_{\mu}\eta_i$.
We find that the former case, as can be seen from the relation  
(\ref{eqn:subtraction}), 
corresponds to the flux-tube which has the quark source.
On the other hands, the latter case, it does not contain 
any information of the quark, which means no terminal,
hence, it cannot provide the physical state like a $q$-$\bar{q}$ system.
If one assumes the existence of the external color-electric source or 
the glueball state as the flux-tube ring\cite{Koma:1999sm}, 
it should be taken into account.
However, since this is not the case which we discuss in this paper,
we assume that there is no singularity in $\partial_{\mu} \eta_i$. 
Then, this term can be absorbed into 
the regular dual gauge field $B_{i\;\mu}^{\rm reg}$ by the
replacement  
$B_{i\;\mu}^{\rm reg} + \partial_{\mu} \eta_i /g'
\to B_{i\;\mu}^{\rm reg}$.
In this case, the flux (\ref{eqn:flux2}) just has the meaning of the 
boundary condition of the regular dual gauge field which should behave as
$B_{i\;\mu}^{\rm reg}\to -B_{i\;\mu}^{\rm sing}$ at infinity, 
where monopoles are condensed.

\section{The static $q$-$\bar{q}$ system}\label{sec:static}
In this section, we consider the static $q$-$\bar{q}$ system.
The quark source is given by the $c$-number current, 
which is typical in the heavy quark system,
\be
{\vec{j}^{\mu}}
\equiv
{\vec{j}^{\mu}}_{j}(x)
=
 \vec{Q}_{j} g^{\mu 0}
\left[
  \delta^{(3)} \left (\bvec{x} - \bvec{a} \right )
- \delta^{(3)} \left (\bvec{x} - \bvec{b} \right )
\right ],
\label{eqn:current}
\ee
where $\vec{Q}_{j} \equiv e \vec{w}_{j}$ is the
Abelian color-electric charge of the quark.
Here, $\bvec{a}$ and $\bvec{b}$ are position vectors
of the quark and the antiquark, respectively, and $\vec{w}_{j}$ is the
weight vector of $SU(3)$ algebra,
$\vec{w}_1= \left (1/2, \sqrt{3}/6 \right ), \vec{w}_2= \left
(-1/2, \sqrt{3}/6 \right ),
 \vec{w}_3= \left (0, -1/\sqrt{3} \right )$.
This vector is nothing but the diagonal component of $\vec{H}=(T_3,T_8)$.
The label $j =1, 2, 3$ can be assigned to the charge
red($R$), blue($B$) and green($G$).
We assume the cylindrical geometry of the system by
taking $\bvec{a} = -(r/2) \bvec{e}_z$, $\bvec{b} = (r/2) \bvec{e}_z$, 
and $n_{\mu} = \bvec{e}_z$, where the distance between the quark 
and the anti-quark is defined by $r$.
In this system, we get an explicit form of the singular dual gauge field
from the relation (\ref{eqn:subtraction}) as
\bea
\bvec{B}_{i}^{\rm sing} = \sqrt{\frac{2}{3}}
 \vec{\epsilon}_{i}\cdot \left [ -
\frac{\vec{Q}_{j}}{4\pi\rho}
\left ( 
 \frac{z+r/2}{\sqrt{\rho^2 +(z+r/2)^2}}
-\frac{z-r/2}{\sqrt{\rho^2 +(z-r/2)^2}}
\right ) \bvec{e}_{\varphi} \right ],
\label{eqn:b-sing-fin}
\eea
where $\varphi$ is the azimuthal angle around the $z$-axis 
and $\rho$ denotes the radial coordinate.
Since the color-electric charges are defined on
the weight vector of $SU(3)$ algebra,  
there arises the relation
\be
\vec{\epsilon}_i \cdot \vec{w}_j = 
-\frac{1}{2} 
\left (
\begin{array}{ccc}
  0 &  1 &  -1 \\
 -1 &  0 &   1 \\
  1 & -1 &   0 
\end{array} 
\right ) 
=
-\frac{1}{2}\sum_{k=1}^3 \epsilon_{ijk}
\equiv 
- \frac{1}{2} m_{i j},
\label{eqn:charge-relation}
\ee
where $m_{i j}$ takes 0 or $\pm$ 1.
The zero of the diagonal component means that one of the 
monopole field is decoupled from the 
system and it does not contribute to the energy 
when we pay attention to the one of the color-electric charge,
since the color-magnetic charge of the monopole field is defined on
the root vector of $SU(3)$ algebra, as $g\vec{\epsilon}_i$.

\par
Here, we investigate the ideal system for the limit $r \to \infty$.
That is,
\bea
\lim_{r \to \infty} \bvec{B}_{i}^{\rm sing} 
=
\sqrt{\frac{2}{3}} \frac{e m_{ij}}{4\pi\rho} \bvec{e}_{\varphi} 
= 
\frac{ m_{ij}}{g'\rho} \bvec{e}_{\varphi},
\label{eqn:b-sing-inf}
\eea
where we have used $eg=4\pi$ and $g'=\sqrt{3/2}g$.
Then, the fields depend only on the radial coordinate,
\be
\phi_i = \phi_i(\rho),\quad
\bvec{B}_{i}^{\rm reg} = B_{i}^{\rm reg}(\rho)\bvec{e}_{\varphi} 
\equiv  
\frac{\tilde B_{i}^{\rm reg}(\rho)}{\rho}\bvec{e}_{\varphi},
\ee
and the field equations (\ref{eqn:feq-1}) and (\ref{eqn:feq-2}) 
are reduced to 
\bea
&&
\frac{d^2\phi_i}{d\rho^2} +\frac{1}{\rho}\frac{d\phi_i}{d\rho}
-
\left ( 
\frac{ g' \tilde B^{\rm reg}_{i} + m_{ij}}
{\rho} \right )^2 \phi_i 
- 2 \lambda \phi_i (\phi_i^2 - v^2) = 0,
\label{eqn:f-eq-cyli-1}
\\
&&\nonumber\\
&&
  \frac{d^2 \tilde B^{\rm reg}_{i}}{d\rho^2} 
- \frac{1}{\rho}\frac{d \tilde B^{\rm reg}_{i}}{d\rho}
-2 g' \left ( g' \tilde B^{\rm reg}_{i} + m_{ij} \right ) \phi_i^2 = 0 ,
\label{eqn:f-eq-cyli-2}
\eea
The string tension can be defined by the energy  
per unit length of the flux-tube,
\bea
\sigma 
= 
2\pi 
\sum_{i=1}^3 
\int_0^{\infty} \rho d\rho
\left [
\frac{1}{2} \left ( \frac{1}{\rho}\frac{d \tilde B^{\rm reg}_{i}}
{d\rho} \right )^2
+
\left ( \frac{d \phi_i}{d \rho} \right)^2
+
\left ( 
\frac{ g' \tilde B^{\rm reg}_{i} +  m_{ij}}{\rho} 
\right )^2 \phi_i^2
+
\lambda ( \phi_i^2 - v^2 )^2
\right ],
\label{eqn:string-tension}
\eea
and we obtain the flux quantization condition,
\be
\Phi_i = - \frac{2 \pi m_{ij}}{g'}.
\label{eqn:flux3}
\ee
The boundary conditions are given by
\bea
&&
\tilde B^{\rm reg}_{i} = 0, \quad \phi_i = 
\left  \{
\begin{array}{cc}
0 & (i\ne j)\\
v & (i = j)
\end{array}
\right.
\quad {\rm as}\quad \rho \to 0, \nonumber\\
&&
\tilde B^{\rm reg}_{i} = - \frac{m_{ij}}{g'},\quad \phi_i = v 
\qquad\qquad\;\;\; {\rm as}\quad  \rho \to \infty.
\label{eqn:boundary-condition}
\eea

\par
Here, we shall confirm the relation (\ref{eqn:subtraction}).
In this cylindrical system, the non-local term can be computed 
explicitly,
\bea
\sqrt{\frac{2}{3}} \vec{\epsilon}_{i} \cdot
\frac{1}{n \cdot \partial}\varepsilon_{\mu \nu \alpha \beta}
n^{\alpha} \vec{j}^{\beta} 
&=&
\sqrt{\frac{2}{3}} \vec{\epsilon}_{i} \cdot
 - \vec{Q}_{j} \delta (x) \delta (y) \bvec{e}_z \quad 
\left (\vec{Q}_j \equiv e \vec{w}_j \right )
\nonumber\\
&=&
\bvec{\nabla} \times 
\left ( \frac{m_{ij}}{g'\rho} \bvec{e}_{\varphi} \right ).
\label{eqn:singular-line}
\eea
As can be seen from this expression, one finds that this term exactly 
cancels with the color-electric field which is originated from the
singular dual gauge field $\bvec{B}_{i}^{\rm sing}$ 
in (\ref{eqn:b-sing-inf}).
It shows that the kinetic term of the dual gauge 
field in the Lagrangian (\ref{eqn:dgl-weyl}) can be written
with no singular field.

\section{Bogomol'nyi limit}\label{sec:Bogomolnyi}
In this section, we discuss the properties of the flux-tube in the 
Bogomol'nyi limit.
Since now we have the same Lagrangian with $U(1)$ gauge symmetry
except only the labels of $i$ and $j$ which classify
the kinds of the monopole and the quark corresponding to $[U(1)]_m^3$
dual gauge symmetry, we can use the same strategy to find the 
Bogomol'nyi limit as given in Ref. \cite{Bogomolny:1976de}.
Thus, we can write the string tension (\ref{eqn:string-tension}) 
exactly in the form,
\bea
\sigma 
&=& 
2\pi \sum_{i=1}^{3}|m_{ij}|v^2
+
2\pi 
\sum_{i=1}^3 
\int_0^{\infty} \rho d\rho
\Biggl [
\frac{1}{2} \left ( \frac{1}{\rho}\frac{d \tilde B^{\rm reg}_{i}}
{d\rho} 
\pm g'( \phi_i^2 - v^2 )
\right )^2 \nonumber\\
&&
+
\left ( \frac{d \phi_i}{d \rho} 
\pm
\left ( g' \tilde B^{\rm reg}_{i} +  m_{ij} \right )
\frac{\phi_i}{\rho}
\right)^2 
+
\frac{1}{2}\left (2 \lambda - g'^2 \right ) ( \phi_i^2 - v^2 )^2
\Biggr ].
\label{eqn:string-tension-2}
\eea
From this expression, we find the Bogomol'nyi limit,
\bea
{g'}^2 = 2\lambda, \quad {\rm or}\quad 3 g^2 = 4 \lambda.
\label{eqn:bogomolnyi-limit}
\eea
In this limit, one find that the string tension is reduced to
\bea
\sigma
&=& 
2 \pi \sum_{i=1}^3 |m_{ij}| v^2 = 4 \pi v^2 ,
\label{eqn:st-exact}
\eea
and the profiles of the dual gauge field and the monopole field
is determined by the first order differential equations,
\bea
&&
\frac{d\phi_i}{d\rho} 
\pm 
\left ( g'\tilde B^{\rm reg}_{i}  + m_{ij} \right )
\frac{\phi_i}{\rho}=0, 
\label{eqn:bogomol-eq-1}\\
&&
\frac{1}{\rho} \frac{d \tilde B^{\rm reg}_{i} }{d\rho}
\pm
g'(\phi_i^2 - v ^2) =0.
\label{eqn:bogomol-eq-2}
\eea
These field equations of cource reproduce the second order 
differential equations (\ref{eqn:f-eq-cyli-1}) and
(\ref{eqn:f-eq-cyli-2}) when the relation 
(\ref{eqn:bogomolnyi-limit}) is satisfied.

\par
Here, to obtain the string tension of the form 
(\ref{eqn:string-tension-2}) and the 
saturated string tension (\ref{eqn:st-exact}), we 
have paid attention to the boundary
conditions of the fields 
(\ref{eqn:boundary-condition})
by taking into account the relation
(\ref{eqn:charge-relation}).
For instance, let us consider the $R$-$\bar{R}$ flux-tube, which is 
given by the label $j=1$.
In this system, the monopole field $\phi_1$ which has the magnetic charge 
$g\vec{\epsilon}_1$ is decoupled from the system, since $\phi_1$
does not feel any singularity of the flux-tube core,
and accordingly, the regular dual gauge field $B_1^{\rm reg}$ is also 
decoupled.
The behavior of the other fields is interesting, $\phi_2$ and $\phi_3$
behaves as the same monopole field, and $B_2^{\rm reg}$ and 
$B_3^{\rm reg}$ provides the $U(1)_{i=2}$ flux-tube 
and $U(1)_{i=3}$ {\em anti} flux-tube due to 
the sign of the $m_{ij}$, which takes $1$ and $-1$, respectively.
Here, both dual gauge fields are related with each other through
the constraint $\sum_{i=1}^3 B_i^{\rm reg}=0$, and
$U(1)_{i=3}$ anti flux-tube can be regarded as 
the $U(1)_{i=2}$ flux-tube, or vise versa.
As a result, these flux-tubes provide the same string tension $2\pi v^2$,
and finally, we get two times of this string tension, $2 \times 2\pi v^2$.
This is caused by the $[U(1)]_m^2$ dual gauge symmetry.
We note that this discussion is the Weyl symmetric, and thus,
the final expression for the string tension (\ref{eqn:st-exact}) 
does not depend 
on kind of the color-electric charges $\vec{Q}_j$.
The profiles of the color-electric field can be obtained by 
solving the first order equations (\ref{eqn:bogomol-eq-1}) and 
(\ref{eqn:bogomol-eq-2}) by taking into account the above discussion
as is discussed in Ref.~\cite{Bogomolny:1976de,deVega:1976mi}.

\par
Let us consider the meaning of (\ref{eqn:bogomolnyi-limit}).
Here, we can define two characteristic scales using three 
parameters in the DGL theory, $g$, $\lambda$ and $v$.
One is the mass of the dual gauge field $m_{B}= \sqrt{2}g'v = \sqrt{3}gv$ 
and the other is the mass of the monopole 
field $m_\chi = 2 \sqrt{\lambda}v$.
These masses are extracted from the Lagrangian (\ref{eqn:dgl-weyl})
by taking into account the dual Higgs mechanism.
Thus, one finds that the Bogomol'nyi limit in the DGL theroy 
(\ref{eqn:bogomolnyi-limit}) is the supersymmetry 
between the dual gauge field and the monopole field.
Since these inverse masses $m_B^{-1}$ and $m_\chi^{-1}$ corresponds to the 
penetration depth of the color-electric field and the coherent length 
of the monopole field, respectively, the Ginzburg-Landau (GL) 
parameter is defined:
\bea
\tilde \kappa \equiv \frac{m_B^{-1}}{m_\chi^{-1}}= \frac{\sqrt{2\lambda}}{g'}
= \frac{2\sqrt{\lambda}}{\sqrt{3}g}.
\eea
Therefore, $\tilde \kappa=1$ is regarded as the Bogomol'nyi limit,
and the vacuum is classified into two types in terms of the 
Bogomol'nyi limit: 
$\tilde \kappa < 1$ belongs to the type-I vacuum and 
$\tilde \kappa > 1$ is the type-II vacuum.

\par
Now, we would like to discuss the interaction between two 
parallel flux-tubes of the same type,  
such as the system $R$-$\bar{R}$ and $R$-$\bar{R}$.
In general, the flux-tubes would interact with each other.
However, in the Bogomol'nyi limit, there is no interaction between them.
This can be understood through an investigtion of the generalized 
string tension for an exotica that the 
color-electric charges are given 
by $n\vec{Q}_j$ and $-n\vec{Q}_j$ for an integer $n$.
In this system, we get the generalized string tension,
\bea
\sigma_n = 4 \pi n v^2,
\label{eqn:gene-st-exact}
\eea
where $m_{ij}$ is simply replaced to $n m_{ij}$.
One finds that the string tension (\ref{eqn:gene-st-exact}) 
is proportional to $n$, which implies that 
the interaction energy is zero.
It is considered that this comes from the balance of propagation range 
of the dual gauge field and the monopole field since $m_B \sim m_\chi$.
In the type-I or in the type-II vacuum, which is 
away from the Bogomol'nyi limit, the interaction range of these fields 
lose its balance, and the flux-tube interaction manifestly appears.
The string tension is not proportional to $n$ any more.
While the attractive force is worked between two parallel flux-tubes in
the type-I vacuum, the flux-tubes repel each other in the type-II vacuum.
Numerical investigations of the interaction between 
two or more parallel flux-tubes of the same type in the DGL theory
are given in Ref. \cite{Koma:1997qq,Ichie:1996jr}.

\par
It is interesting to investigate what happens if 
two parallel flux-tubes of different types 
are placed at a certain distance\cite{Suganuma:1995ij}.
Here, according to the $[U(1)]_m^3$ dual gauge symmetry, there appear
three different types of the flux-tube, such as given by
$R$-$\bar{R}$, $B$-$\bar{B}$, and $G$-$\bar{G}$, so that 
these interactions seem to be very complicated.
However, now the system has remarkable aspects 
owing to the Weyl symmetry.
For instance, let us consider the interaction 
between $R$-$\bar{R}$ and $B$-$\bar{B}$.
We find that the interaction between them is attractive, 
since if we suppose that these flux-tubes are unified into 
one flux-tube, it becomes $\bar{G}$-$G$ (See the relation 
(\ref{eqn:charge-relation})). 
It means that the energy of the system after unification is
reduced into a half of the initial one. The same interaction 
property would be observed in
the process,
$B$-$\bar{B}$ + $G$-$\bar{G}$ $\to$ $\bar{R}$-$R$ and 
$G$-$\bar{G}$ + $R$-$\bar{R}$ $\to$ $\bar{B}$-$B$.
These investigations show that
if we pay attention to the Weyl symmetry,
we can easily obtain qualitative information about 
the flux-tube interaction.

\section{Conclusion}
We have studied the flux-tube solution in the DGL theory
in the Bogomol'nyi limit by using the manifestly Weyl invariant form 
of the DGL Lagrangian. 
Here, the original dual gauge symmetry $[U(1)]_m^2$ is 
extended to $[U(1)]_m^3$.
This replacement makes the further manipulation of the 
Lagrangian analogous to the $U(1)$ case.
We have found that the Bogomol'nyi limit is given by $3 g^2 = 4 \lambda$, 
and the string tension is calculated as $\sigma_n = 4\pi n v^2$
for a $q$-$\bar{q}$ pair with the charge $nQ_j$ and $-nQ_j$ in the 
both ends.
In this limit, the mass of the dual gauge field
and the monopole field becomes {\em exactly} the same.
It should be noted that we could see the same relation with $U(1)$ Abelian 
Higgs theory except for three different types of the flux-tube.
To summarize, the very similar properties with the ANO vortex
in the Abelian Higgs theory is observed when we see the single 
flux-tube in the DGL theory, and the flux-tube solution can 
be easily obtained if we pay attention to the Weyl symmetry 
in the the DGL theory.

\par
Finally, we would like to mention about the relation between 
the work in Ref. \cite{Chernodub:1999xi} and our study. 
If we replace the monopole field and the parameters that they have used 
as $\chi \to \sqrt{2}\chi$, $\eta \to \sqrt{2}v$, and 
$\lambda \to \lambda/4$, we get the same framework at the starting 
point, and the Bogomol'nyi limit is replaced as 
$3g^2 = 16\lambda \to 3g^2 =4\lambda$.
The idea of the extension of the dual gauge symmetry based on the Weyl
symmetry in our case, however, seems to be simple to reach the final 
expression on the string tension, which can be applied 
to the $[U(1)]^{N-1}$ dual Abelian Higgs theroy reduced from 
the $SU(N)$ gluodynamics, straightforwardly.

\section*{Acknowledgment}
The authors are grateful to H. Suganuma for fruitful discussions and in 
particular stressing the importance of the Weyl symmetry in the DGL theory.
We also acknowledge M. I. Polikarpov to inform us of 
the work of Ref. \cite{Chernodub:1999xi}, which is closely related to 
our work.


\end{document}